\documentstyle{l-aa}

\begin{document}

\newcommand{\equ}{\begin{equation}}
\newcommand{\eequ}{\end{equation}}
\newcommand{\arry}{\begin{eqnarray}}
\newcommand{\earry}{\end{eqnarray}}
\newcommand{\BF}{\begin{figure}}
\newcommand{\EF}{\end{figure}}
\newcommand{\BI}{\begin{itemize}}
\newcommand{\EI}{\end{itemize}}
\newcommand{\BE}{\begin{enumerate}}
\newcommand{\EE}{\end{enumerate}}
\newcommand{\dis}{\displaystyle}
\newcommand{\BC}{\begin{center}}
\newcommand{\EC}{\end{center}}
\newcommand{\BL}{\begin{flushleft}}
\newcommand{\EL}{\end{flushleft}}
\newcommand{\BTA}{\begin{table}}
\newcommand{\ETA}{\end{table}}
\newcommand{\BT}{\begin{tabbing}}
\newcommand{\ET}{\end{tabbing}}
\newcommand{\TAB}{\begin{tabular}}
\newcommand{\ETAB}{\end{tabular}}
\newcommand{\BD}{\begin{description}}
\newcommand{\ED}{\end{description}}

\newcommand{\ApJ}{{ApJ }}
\newcommand{\AsA}{{A\&A }}
\newcommand{\AsJ}{{AJ }}
\newcommand{\Mn}{{MNRAS }} 
\newcommand{\Asp}{{\em Astrophys. Space Sci.}}
\newcommand{\ApJS}{{ApJS }}
\newcommand{\AsAS}{{A\&AS }}
\newcommand{\JQSRT}{{\em J. Quant. Spectros. Radiat. Transfer}}
\newcommand{\ARAS}{{\em Ann. Rev. Astr. Ap.}}
\newcommand{\Via}{{\em Vistas in Astronomy}}
\newcommand{\ApJL}{{\em Astrophys. J. Lett.}}
\newcommand{\Na}{{\em Nature}}
\newcommand{\AnA}{{\em Ann. d'Ap.}} 
\newcommand{\PhR}{{\em Phys. Rev.}}
\newcommand{\Nse}{{\em Nucl. Sci. Engng.}}
\newcommand{\Msait}{{Mem. Soc. Astron. Ital. }}

\newcommand{\Spe}{{Spectral Evolution of Galaxies}}
\newcommand{\RT}{{\em Radiative Transfer}}
\newcommand{\Ntt}{{\em Neutron Transport Theory}}
\newcommand{\Ppim}{{\em Physical Processes in the Intestellar Medium}}
\newcommand{\Iaunu}{{\em IAU Symp.\ 1991}}
\newcommand{\Iauoq}{{\em IAU Symp.\ No.\ 108}}
\newcommand{\Rmcr}{{\em Recent developments of Magellanic Cloud Research}}
\newcommand{\Sn}{{\em Stellar Nucleosynthesis}}
\newcommand{\Iauon}{{\em IAU Symp.\ No.\ 135 1989, }}
\newcommand{\NIM}{{\em Nebulae and Interstellar Matter}}
\newcommand{\Iauno}{{\em IAU Symp.\ No.\  1990, The Galactic and Extragalactic}}
\newcommand{\Lssp}{{\em Light Scattering by Small Particles}}
\newcommand{\AGAS}{{\em Astrophysics of Gaseous Nebulae and Active Galactic
Nuclei}}
\newcommand{\PGAS}{{\em Physics of Thermal Gaseous Nebulae}}
\newcommand{\ND}{{\em Numerical Data and Functional Relationships in Science
and Technology}}
\newcommand{\nrsc}{{\em New Results on Standard Candles}}
\newcommand{\RMAA}{{\em Rev. Mex. A.A.}}
\newcommand{\GEO}{{\em Geochim. et Cosmochim. Acta}} 
\newcommand{\Pasp}{{PASP }}

\thesaurus{3(11.03.1; 11.05.2; 11.19.3; 11.19.5; 09.08.1)}

\title{Starbursts and the Butcher-Oemler effect in galaxy clusters}

\author{B.M.~Poggianti \inst{1,2}
\and G.~Barbaro
\inst{1} }

\offprints{B.M.~Poggianti (Kapteyn Instituut)}

\institute{Dipartimento di Astronomia, vicolo dell'Osservatorio 5, 35122 
Padova, Italy, barbaro@astrpd.pd.astro.it
\and Kapteyn Instituut, P.O. Box 800, 9700 AV Groningen, The Netherlands,
bianca@astro.rug.nl}

\date{Received; accepted}

\maketitle
%
%
%
%
%
%
%
%
%
%
%
%
%
%

\begin{abstract}
In order to explain the spectroscopic observations of most of the galaxies
in intermediate redshift clusters, bursts of star formation superimposed 
to the traditional scenario of galactic evolution are needed.

The analysis of spectral lines and colours by means of an evolutionary
synthesis model, including both the stellar contribution and the emission of
the ionized gas, allows in most of the cases the determination of the time
elapsed since the end of the burst and the fraction of galactic mass involved
in it. In the four clusters considered (AC103, AC114, AC118 at $\rm z=0.31$ and
Cl1358+6245 at  $\rm z=0.33$), the theoretical analysis demonstrates that the
bursts affect substantial galactic mass fractions, typically 30 \% or more. 

The observations can be equally well reproduced by either elliptical+burst 
models or by spiral+burst models in which the star formation is truncated 
at the end of the burst. A way to determine the galactic original type is 
suggested.


\end{abstract}   


\section{Introduction}

Butcher \& Oemler  (1978, 1984) first showed that the fraction of blue galaxies
in rich compact clusters at intermediate redshift ($\rm z < 0.55$) is larger
than in similar local ones. The galaxies responsible for this phenomenon,
commonly referred to as Butcher-Oemler (BO) effect, were tentatively identified
by Butcher and Oemler as normal spirals, subsequently evolving into S0 galaxies
due to the depletion of gas in star forming processes. Further works have
confirmed the reality of this phenomenon (Couch \& Newell 1984; Dressler et al.
1985; Lavery \& Henry 1986, 1988, 1994; Couch \& Sharples 1987 (CS87); MacLaren
et al. 1988; Mellier et al. 1988; Soucail et al. 1988; Dressler \& Gunn 1992;
Rakos \& Schombert 1995).  Especially spectroscopic observations have made a
fundamental contribution to its understanding by achieving three main
conclusions: 

a) the membership of the majority of blue objects has been demonstrated
through the determination of their redshift; 

b) only in some cases  the blue galaxies of the spectroscopic sample have
spectra of normal spirals; generally the blue component exhibits four different
types of spectra: 1) spectra with emission lines (in particular the line
[OII]$\lambda$3727) and with absorption Balmer lines of moderate intensity:
these spectra are similar to those of local spirals; 2) spectra with strong
emission lines and with Balmer absorption lines almost filled in by the
emission of the ionized gas in HII regions: these features typify starburst
galaxies; 3) spectra with emission lines characteristic of galaxies hosting an
AGN; 4) spectra lacking emission lines and with strong absorption Balmer lines:
these spectra do not correspond to any normal galaxy of the Hubble sequence.
The relative fraction of blue galaxies belonging to the four types is different
from cluster to cluster: in some of them the AGNs are in excess (3C295, $\rm
z=0.46$; Dressler and Gunn 1983) in others the blue population is dominated by
normal spirals (0024+1654, $\rm z=0.39$; Dressler et al. 1985). However in most
cases the largest fraction consists of type 4) spectra. 

c) also several ``red'' galaxies, not responsible for the BO effect, display
unexpected spectral features: the absorption Balmer lines are considerably
stronger than in the corresponding local galaxies: the normal ellipticals. Such
objects have been called E+A galaxies (Gunn \& Dressler 1988) since their
spectra can be reproduced by summing up the spectrum of an elliptical galaxy
and the spectrum of a stellar population dominated by A type stars. Similar
anomalies have been also found by means of multicolour intermediate-band
photometry. Ellis et al. (1985) and MacLaren et al. (1988) found in clusters
0016+16 ($\rm z=0.54$) and Abell 370 ($\rm z=0.37$) that some galaxies, whose
optical colours are similar to normal ellipticals, exhibit an UV excess. Other
elements suggesting an anomalous star forming activity are derived from
infrared colours (Lilly 1987; Arag\'on-Salamanca et al. 1991). 

Furthermore, high spatial resolution HST observations (Couch et al. 1994
(CESS94); Dressler et al. 1994; Wirth et al. 1994; Oemler et al. 1995) have
provided informations about the morphological appearance of the cluster
galaxies and about the high occurrence of mergers/interactions. 

Observations suggest that a large fraction of galaxies in rich clusters with
intermediate redshift are involved in some anomalous type of star formation, in
progress or ended in a relatively recent past, unexpected on the basis of the
present knowledge of galaxy evolution. The hypothesis that bursts of star
formation could be responsible for all the summarized phenomenology has been
analysed in a series of papers (Dressler \& Gunn 1982, 1983; Couch \& Sharples
1987; MacLaren et al. 1988; Pickles \& van der Kruit 1988, 1991; Newberry et
al. 1990; Jablonka et al. 1990; Arag\'on-Salamanca et al. 1991; Charlot \& Silk
1994; Barbaro \& Poggianti 1995a; Poggianti \& Barbaro 1994, 1995; Jablonka \&
Alloin 1995; Belloni et al. 1995). The alternative possibility of truncation of
the star formation in late type galaxies has been also explored in some of the
previous papers but such hypothesis seems unable to account for all the
variety of the observed facts. 

In most cases the theoretical analyses consider only few observable quantities
and, in particular, no model presents synthetic estimates of the emission lines
due to the gas ionized by hot young stars in HII regions (only CS87 consider
the emission-filling of the $\rm H\delta$ line, the emission of the gas being
extracted from the spectrum of a spiral galaxy). Consequently the analysis has
been mainly restricted to post-starburst objects. Moreover the clusters have
been generally studied with different observational techniques (photometric and
spectroscopic) and this hinders the intercomparison of the results. 

The aim of this work is to derive a theoretical picture which allows the
modelling of spectral characteristics and colour indexes and at the same time
can be a support for correlating the observational quantities. The tool of this
analysis is a spectrophotometric model able to synthesize the integrated
spectrum of a galaxy by taking into account  the stellar contribution and  the
emission of the ionized gas. The model allows the computation of the Balmer
lines (including the stellar absorption and the gaseous emission) and the
principal emission metallic lines such as [OII]$\lambda$3727. This analysis can
shed some light on the nature of the galaxies connected to the BO effect, their
original morphological type, their further evolution and therefore their local
equivalents. Moreover, since the idea of the bursts of star formation presently
seems to be the most reasonable working hypothesis, we want to make a
systematic analysis of all the parameters to it connected, namely the rate of
star formation, the duration of the burst and the amount of gas involved. 

In this work the analysis is done for a distance and an age corresponding to
$\rm z=0.31$ which is the redshift of three clusters studied in detail by CS87
and also the approximate redshift of the cluster Cl1358+6245 (Fabricant et al.
1991) to which our considerations will be applied. 

\section{The spectrophotometric model}

While a detailed description of the spectrophotometric model is deferred to a
further paper (Barbaro \& Poggianti 1996, Paper II) here we present only the
basic elements which can be useful to critically understand the conclusions. 

- The stellar spectrum is computed with an improved version of Barbaro and
Olivi's model (1986, 1989) which includes all the stellar  advanced
evolutionary phases, up to the post-AGB, and takes into account stellar
population of different metallicities according to a simple model of chemical
evolution. Changes are introduced in the treatment of stellar atmospheres by:
a) using Kurucz's new models in the 1993 version, b) employing the library of
Jacoby et al. (1984) when a larger resolution was required, as in the case of
the synthesis of the Balmer lines (it must be remarked that such spectra have
solar metallicity and therefore they should not be proper for modelling
metallicity-dependent features in metal-poor populations), c) extending the
computation to the infrared regions with Kurucz's models for stars with 
$T_{\rm eff}>5500$ K and the library of observed spectra of Lan\c{c}on-Rocca
Volmerange (1992, LRV) for stars with lower temperatures. 

- The gaseous non thermal emission and the dust emission are ignored. Also the
dust extinction, which is important in late type galaxies, is not considered:
we have preferred to omit the correction instead of using the standard
extinction curve of our galaxy, since we completely ignore the dust properties
in distant galaxies. 

- The emission of the ionized gas is derived by modelling the galaxy spectrum 
as a superposition of the spectra of HII regions, each of them being excited 
by a single stellar cluster. It is assumed that such HII regions do not 
overlap, which is reasonable provided that the star formation rate is not 
exceedingly high.        

- There must be coherence between the stellar and the gaseous spectrum
in the sense that the UV photon flux must be derived from the same
stars responsible for the stellar emission; this requirement is satisfied 
by fixing the absolute V magnitude and corresponds to determine the
values of the normalization constants in the stellar distribution function.

- In deriving the flux of the metallic lines, the integrated spectrum is
approximated with the spectrum of a single equivalent star with appropriate
temperature and luminosity equal to the integrated luminosity of the cluster.
The validity of this approximation has been confirmed through a test performed
with the photoionization code Cloudy (Ferland 1991), by comparing the results
of the integrated spectrum with those of the equivalent star. It was concluded
that the results are correct within the uncertainties intrinsic to the
evaluation of the lines intensities. These have been derived from Stasinska's
models (1990); the inspection of Stasinska's tables shows that the ratios of
the lines intensities are generally functions both of the temperature and of
the luminosity of the exciting source. In particular this not true for the
ratios of the Balmer lines which practically depend only on the temperature. 

- The first eight lines of the Balmer series, the $\rm Ly \, \alpha$ line and
several metallic lines have been synthesized; only the $\rm H\delta$ and the
[OII]$\lambda$3727 lines have been considered in the present work. 

From the computed spectra several colour indexes in different photometric
systems  have been derived: the UBVRIJHK system of Johnson (Johnson's UBVRI and
JHK from Bessell \& Brett 1988), the (U-685) and (418-685) colours of the
Durham system (Couch et al. 1983), the $\rm (B_{J}-R_{F})$ used by CS87 (Couch
and Newell 1980), the Kron-Cousin system (Bessell 1986), the \sl gri \rm system
(Schneider et al. 1983), the (1550 -V) colour, originally defined by Burstein
et al. (1988). As in the following we are referring to galaxies with $\rm
z=0.32$ in the observer's frame, the response of the 1550 filter at wavelengths
shorter than the lower limit of the computed spectrum (1012 \AA $\,$ in the
rest frame and 1335 \AA $\,$ in the observer's frame) has been set equal to 0;
this however does not affect at least the qualitative conclusions of our
analysis. Moreover the following spectral features have been synthesized: $\rm
D_{4000}$, EW([OII]) of the line at $\lambda=3727$ \AA $\,$ and the intensities
of the $\rm H \alpha$, $\rm H \beta$, $\rm H \delta$ lines (including
absorption and emission components). 

Although observations for these quantities are not yet available for the
clusters here considered, results concerning $\rm D_{4000}$ are included
because of the equivalence between this index and the colour $\rm
(B_{J}-R_{F})$ at $z=0.31$; (U-685) is considered because such colour is
available for another cluster (Abell 370, Arag\'on-Salamanca et al. 1991) and a
theoretical analysis can be useful in order to determine the connection between
the $\rm H \delta$ strong galaxies and the UV-excess phenomenon observed in
other clusters. 

Model results for the different galactic types are shown in Table 1; the case 
of a constantly increasing SFR (Extreme case) has also been considered. 

\BTA

\caption[]{Model results for different normal Hubble types of age 10.4 Gyr
and z=0.32. Equivalent width values are given in the rest frame, while the other
quantities are referred to  the observer's frame. The \sl Ext. \rm model
represents
a galaxy whose SFR increases steadily during the evolution}

\begin{flushleft}

\TAB{lcrccr}          
\hline
\noalign{\smallskip}
Type & $\rm D_{4000}$ & $\rm (1550-V)$ & $\rm (B_J-R_F)$ & EW($\rm H\delta$) & EW([OII]) \\
\noalign{\smallskip}
\hline
\noalign{\smallskip}
E    &  2.15 &  4.45    & 2.67      &   0.7           & 0.1 \\
Sa   &  1.67 &  -0.87   & 2.10      &   1.2           & 10.0 \\
Sb   &  1.55 &  -1.28   & 1.91      &   1.2           & 13.4 \\
Sc   &  1.44 &  -1.63   & 1.69      &   1.4           & 16.9 \\
Sd   &  1.37 &  -1.79   & 1.55      &   1.7           & 17.6 \\
Ext. &  1.28 &  -2.12   & 1.30      &   1.3           & 21.8 \\
\noalign{\smallskip}
\hline
\ETAB

\end{flushleft}

\ETA

\section{Models with bursts of star formation}

In this and in the following sections we discuss a set of models in which a
burst of star formation is added to a model galaxy of a given morphological
type. We define a ``starburst'' any normal  galaxy in which an episode of star
formation is superimposed, regardless of its intensity. 

The aim of this analysis is to derive theoretical relations between burst
parameters and observable quantities. For sake of simplicity the star formation
rate during the burst is assumed constant, then the burst is completely
characterized by two parameters: 1) the intensity, defined as the ratio
$\psi_{1}/\psi_{o}$ of the star formation rate during the burst to the initial
SFR, 2) the length $\Delta t$. Another important parameter is the time elapsed
since the end of the burst $\tau=T-T_{f}$ where $T$ is the age of the model and
$T_{f}$ is the time at which the burst is ended. 

From the previous quantities the fraction of galactic mass that goes into stars
during the burst  can be derived:

$
\Delta g \propto (\psi_{1}/\psi_{o}) \Delta t
$ 

In the following, unless otherwise stated, $\Delta t$ and $\tau$ are given in
million year. 

The range of the parameters investigated is wide. 1) the
existing observations are barely enlightening on the burst length: local 
starbursts and mergers seem to suggest timescales larger than usually
considered, of the order of Gyr. The following values have been explored:
$\Delta t=1,10,100,500,1000$ Myr. 2) We want to consider both the case in which
the mass involved in the burst is comparable with the galaxy mass (which 
should mimic a merging episode) and the case in which the gas consumed is 
internal to the galaxy, in which case its mass is relatively small in respect
to the galaxy mass. Consequently the burst intensity is chosen in order
to account for a mass of stars formed equal to  $10^{-6}-100$ \% of the 
total mass. 3) The burst evolution is followed from its beginning up 
to 2 Gyr after its end with the care of analysing in detail the phases  
during which some observable changes more quickly and therefore we have 
considered the following times: $\tau = 0,1,10,30,100,500,1000,2000$ Myr.

An age of 10.4 Gyr, corresponding to a redshift of 0.32, has been adopted for
the  models. The relation between redshift and age has been derived from a
cosmological model of a universe energetically dominated from non-relativistic
matter with  $q_0=0.225$ and $H_0=50 \rm \, Km \, s^{-1} \, Mpc^{-1}$, in which
the age of the universe is 15 Gyr. The colours derived by convolution of the
model spectrum with the filter response functions are referred to the
observer's frame; however the equivalent widths are computed in the rest frame.
The usual convention has been adopted that the equivalent width of the $\rm H
\delta$ line is positive for the absorption line and negative for the emission
line. 

\section{Bursts in elliptical galaxies}

The star formation rate (SFR) of the underlying galaxy is described by an
exponential function with timescale ${\tau}_G$ of 1 Gyr: $\psi=\psi_{o}
exp(-t/{\tau}_G)$; the average metal abundance is solar. 
\begin{figure}
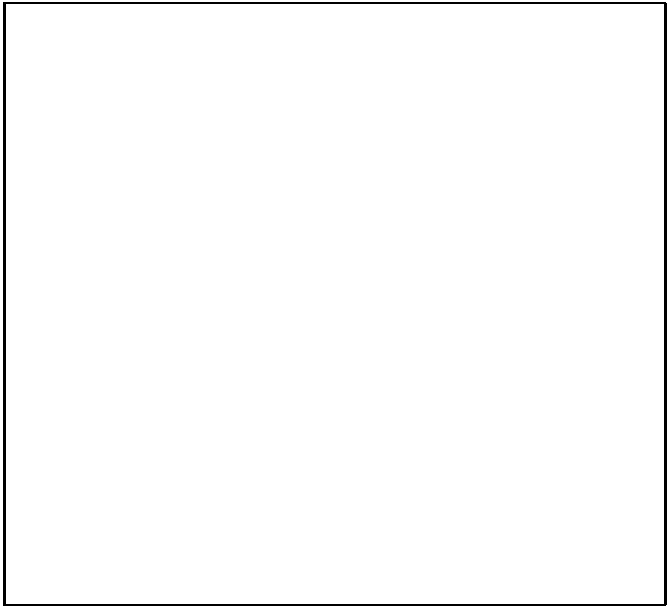

\picplace{8cm}
\caption[]{Model spectra of an unperturbed elliptical 
(top spectrum) and an
elliptical with a burst at three different stages of the evolution (from top to
bottom): when the burst is still active; for $\tau = 10$ Myr; for $\tau = 0.5$
Gyr. All the models are computed for an age of 10.4 Gyr. Fluxes are in
arbitrary units}
\end{figure}

In Fig. 1 the evolution of the spectrum is presented when a burst with $\Delta
t=1$ Gyr and $\Delta g=0.05$ is added to an elliptical galaxy. This case could
be considered as representative of a general situation. The metallic lines have
been computed, but are not shown because they would require a change in the
scale and moreover they are not relevant for the purpose of this figure. From
top to bottom are presented: 1) the spectrum of the unperturbed galaxy, 2) the
spectrum just at the end of the burst: the  $\rm H\alpha$, $\rm H\beta$
emission lines are evident while the $\rm H\delta$ line is emission-filled. The
equivalent width of the [OII]$\lambda$3727 line is equal to 23 \AA. From the
shape of the continuum and the intensity of the lines this spectrum would be
classified as starburst, 3) the spectrum 10 Myr after the end of the burst; the
$\rm H\delta$ line with  EW $\simeq$3.5 \AA $\,$ is clearly evident while the
emission lines are now fainter (EW([OII])=3), because the remaining hot young
stars, although numerous, contribute to the UV flux much less than the massive
stars already evolved out of the main sequence.  With these features an
observed spectrum would be classified as ``blue $\rm H\delta$ strong'', 4) the
spectrum 0.5 Gyr after the end: the only signatures of the burst are the strong
absorption Balmer lines, while emission lines are absent. An object with such a
spectrum would be considered as ``red $\rm H\delta$ strong''.

The time evolution of some observables after the end of the burst is shown in
Fig. 2 for models with $\Delta t=500$ and different values of $\psi_{1}/
\psi_{o}$. In panel a) the behaviour of the equivalent width of the $\rm H
\delta$ line is presented: ``strong'' values (EW $\geq3\,$ \AA), following the
terminology of CS87, are reached only if the mass used in the burst is larger
than  2.5 \%  of the total mass of the galaxy: the curve with full circles
corresponds to this limit and this result is independent of the particular 
choice of  $\Delta t$. The length of the ``strong'' phase is a function of
$\Delta g$ and varies between 100 and 1500 Myr: this last value is an upper
limit since it corresponds to the lifetime of the stars responsible for the
maximum absorption. The time required to reach the maximum is of the order of
10 Myr, independent of the mass implied, and is the time necessary for the
emission of the gas to become negligible compared with the absorption. 

\begin{figure}
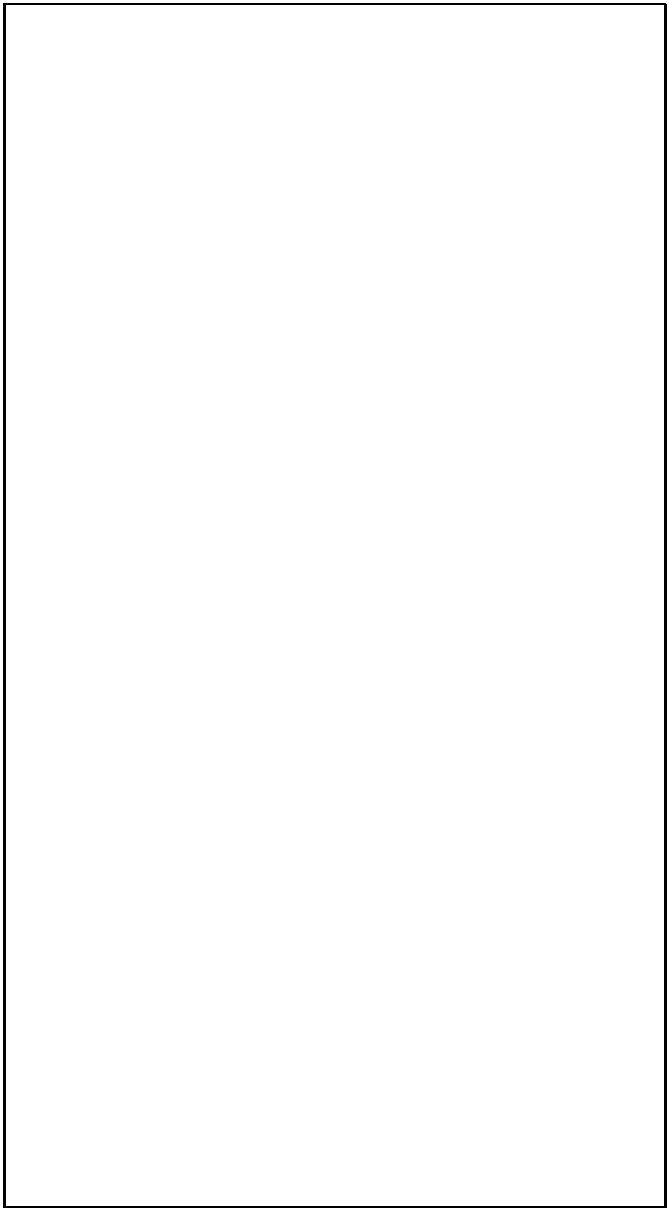

\picplace{16cm}
\caption[]{Time evolution after the end of the burst of: 
\bf a \rm
EW($\rm H\delta$) (\AA), \bf b \rm $\rm (B_J-R_F)$, \bf c \rm $\rm (U-685)$,
\bf d \rm $\rm (1550-V)$. The time $\tau$ is expressed in Myr. Lines connect
models of ellipticals with bursts of $\Delta t = 500$ Myr and $\Delta g =
0.0005$ (empty circles), $\Delta g = 0.005$ (full squares), $\Delta g = 0.025$
(full circles) and $\Delta g = 0.15$ (empty squares)}
\end{figure}

After the end of the strongest bursts ($\Delta g >0.5 $) for the first millions
of years the  $\rm H\delta$ line is in emission (according to the adopted
convention the EW is negative). Such behaviour does not appear in any of the
models with a normal star formation rate (without burst) and therefore is an
unambiguous symptom of a burst still in progress or ended by less than 5
million years. 

Also the [OII]$\lambda 3727$ line evolves quickly, being associated with  those
stars able to ionize the gas; the phase in which such line can be revealed
(when EW $>1 \,$ \AA) is restricted to the burst length and a period of 10 Myr
after its end. There is a short time interval during which the presence of the 
[OII]$\lambda$3727 emission line and of the $\rm H\delta$ absorption line are
detectable on the spectrum. Of course the probability of catching such an event
is small. 

The other emission features behave in a similar way; the $\rm H\alpha$ line
persists in emission for a smaller period than the  [OII]$\lambda$3727 line due
to its superposition with the stellar absorption. The $\rm H\beta$  emission is
fainter for the same reason and in absorption it shows the same behaviour as
the $\rm H \delta$ line. 

In Fig. 2b and 2c the evolution of the colour indexes $\rm (B_{J}-R_{F})$ and
$\rm (U-685)$ are presented for the same models of Fig. 2a. In Fig. 2b the
lower horizontal line marks the limit between blue and red galaxies, according
to the definition of CS87, while the upper one sets the limit beyond which the
bursting galaxy cannot be distinguished from a normal elliptical. The
horizontal line in Fig. 2c separates normal ellipticals from those with UV
excess, the last ones being defined as those having (U-685) bluer by at least
of 0.2 mag. than the average of ellipticals. In both diagrams the colours tend
asymptotically to those of the unperturbed elliptical model. Also for these
colours the time interval during wich they are significantly different from
those of the normal  elliptical increases with $\Delta g$ and can last also 1.5
- 2 Gyr as was found for the $\rm H\delta$ line. 

The other colours that have been synthesized behave in a similar way. Their
sensitivity to the presence of young stars increases with the decrease of the
effective wavelength of the bluer filter; therefore a weak burst affects the
(U-B) and (B-V) colours but is unable of modifying substantially  for instance
the (R-I) colour. In particular the (1550-V) colour index, whose evolution is
described by the diagram of Fig. 2d, is able to reveal a star forming activity
even if the amount of  mass implied is very small (again the horizontal line
represents the colour of the unperturbed elliptical. 

The general features of the evolution of the index $\rm D_{4000}$,
defined as in Bruzual (1983), are similar to those of the ($\rm B_J - R_F$) 
colour; the horizontal line indicates the position of the unperturbed
elliptical. Models show that, at the adopted redshift, the following
relation exists:  $\rm D_{4000} =0.16 \cdot (\rm B_J - R_F)+1$.

The $\rm H\delta$-$\rm (B_J-R_F)$ diagram may  give a positive contribution to
the understanding of the behaviour of starburst galaxies and particularly of
E+A objects.  The evolution in this diagram of the model used in Fig. 1 (solid
line)  and of a second model with $\Delta g=0.15$ and  $\Delta t=500$ Myr
(dashed line) is shown in Fig. 3. The vertical segment and the horizontal 
line mark respectively the boundary between ``blue'' and ``red'' objects and 
between $\rm H\delta$-strong (EW $ \ge 2.5$ \AA $\,$ in this case) and $\rm
H\delta$-weak ones. Numbers give the time elapsed since the end of the burst
(in Myr). The models of an unperturbed elliptical and that corresponding to the
end of the burst are also presented. The discussion of this diagram will be
postponed to section 7 where the theoretical data are compared with the
observations of three clusters. 

\begin{figure}
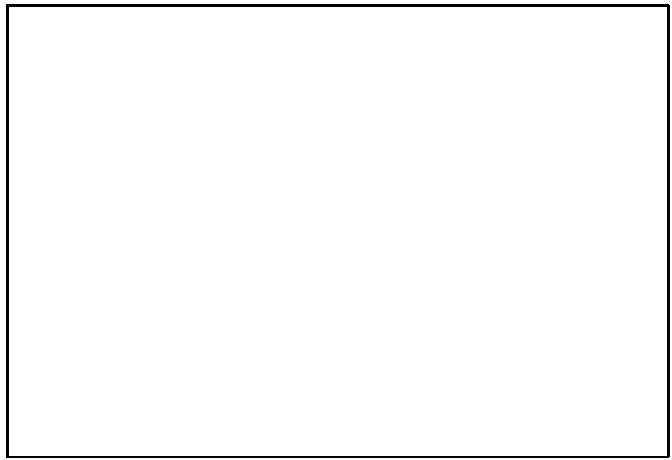

\picplace{6cm}
\caption[]{Time evolution of two ellipticals with burst models 
in the
diagram $\rm H\delta$ versus $\rm (B_J-R_F)$. Numbers give the time elapsed
since the end of the burst (in $10^6$ yr). The empty square marks the position
of the unperturbed elliptical model; the two stars represent the positions of
the $\tau = 0$ burst models. The vertical segment and the horizontal line mark
respectively the division between the ``blue''-``red'' and the ``$\rm H \delta$
strong''- ``$\rm H \delta$ weak'' regions}
\end{figure}

For a large part of its evolution the equivalent width of $\rm H\delta$ is
constant and corresponds to its maximum (Fig. 3).  Moreover this maximum is
joined to the absence of the  [OII]$\lambda$3727  line,\ and to colour indexes
not too red: from all these features it is easy to recognize when the observed
value of $\rm  EW(H\delta)$ is at its maximum. 

Generally the evolution of the observable quantities depends separately on both
$\Delta t$ and $\psi_{1}/\psi_{0}$: this is the case, for instance, of
[OII]$\lambda$3727  and other spectral features and all the colours. Only  the
maximum value of the equivalent width  $\rm EW(H\delta)_{max}$ is correlated
with $\Delta g$ in a simple way. In fact the $\rm EW(H\delta)$ of a single
stellar population is high for a long period of time, up to an age of 1.5-2
Gyr. Therefore if $\Delta t$ is lower than such values all the stellar
populations born during the burst contribute in the same way, approximately as
if they would have the same age and $\rm EW(H\delta)_{max}$ is determined only
by the total fraction of stars formed during the burst. 
$\rm EW(H\delta)_{max}$, which is generally reached about 100 Myr after the end
of the burst, is plotted as a function of $\Delta g$ in Fig. 4 (open circles 
are of concern at present), for different combinations of $\Delta t$ and
(${\psi_{1}}/{\psi_{0}}$). 
\begin{figure}
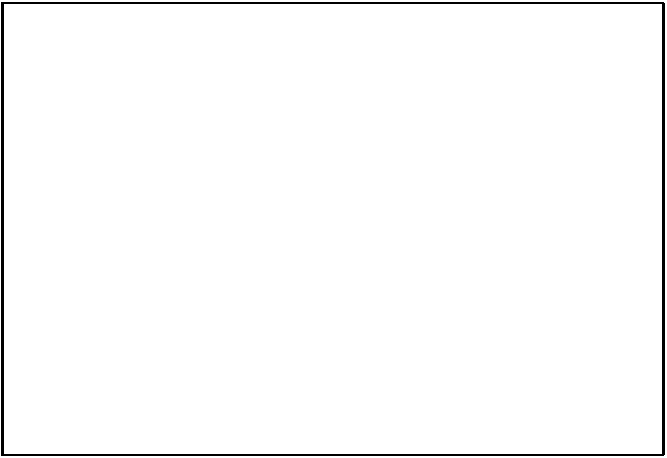

\picplace{6cm}
\caption[]{Dependence of the maximum EW($\rm H\delta$) on 
$\Delta g$ models
of ellipticals with bursts (empty circles)  compared to  truncated spirals with
bursts of different Hubble types (filled circles)}
\end{figure}
Therefore $\rm EW(H\delta)$ can yield an estimation of the amount of mass
involved in the burst. Similar conclusions cannot be reached from EW([OII]) or
$\rm (B_J-R_F)$. $\rm EW(H\delta)_{max}$ is saturated when $\Delta g \geq 0.20
$, that is, when the burst is very strong, it becomes insensitive to the amount
of mass involved and in this case only a lower limit to $\Delta g$ can be set. 

\begin{figure}
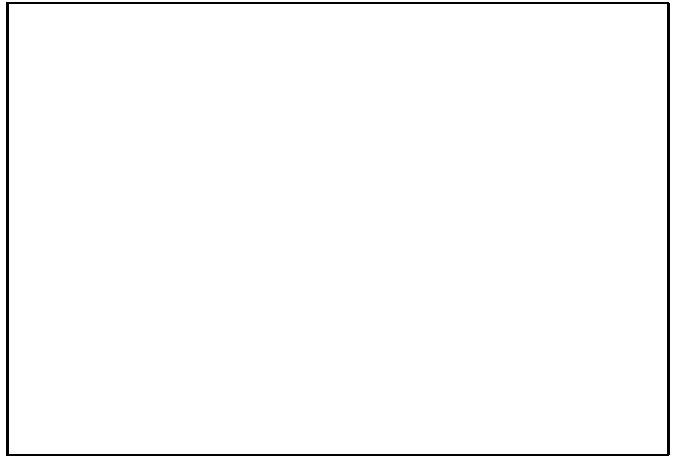

\picplace{6cm}
\caption[]{EW([OII]) (\AA) as a function of $\Delta t$ 
when the burst is
still active ($\tau=0$). Symbols are: $\psi_{1}/\psi_{o}=0.001$ (empty
circles), $\psi_{1}/\psi_{o}=0.01$ (full squares), $\psi_{1}/\psi_{o}=0.05$
(full circles), $\psi_{1}/\psi_{o}=0.3$ (empty squares). The intervals in
$\Delta t$ are: 1,10,100,500,1000 Myr (1,10,100,260, 400,500,670,850,1000 for
the model of highest intensity represented by empty squares)}
\end{figure}

Unlike the line flux, the equivalent width of the  [OII]$\lambda$3727  line is
not proportional to the number of massive stars formed in the last $10^{7}$ yr.
In Fig. 5 EW([OII]) is plotted as a function of $\Delta t$ for different
values of the intensity $\psi_{1}/\psi_{0}$. Provided the intensity is
sufficiently high and the length of the burst is large enough, the equivalent
width decreases as the length increases. The reason for this behaviour is that,
while the models with the same intensity (same symbol) have the same line flux
(the number of ionizing stars being the same), the continuum flux at
$\lambda=3727 \, $ \AA $\,$ is progressively increasing with $\Delta t$, its
contribution arising also from stars with lifetime larger than those
responsible for the photoionization, and therefore the equivalent width
decreases. A similar behaviour is found also in the other emission features, as
for instance in $\rm H\delta$ for $\tau = 0$ which, for bursts strong enough,
is emission-dominated  and in the colours.

\section{Bursts in spiral galaxies}

When considering bursts in spiral galaxies two possibilities can arise:
a) at the end of the burst the SFR can reach again the value typical of
its morphological type or b) the star formation is truncated.
Case a), referred to as ``spirals with bursts'', concerns the situation in which
the amount of material involved in the burst is only a small fraction of the
available gas or is provided by the environment. In any case it is assumed
that the effects of the burst do not hinder any further star formation.
Case b) in which, on the contrary, almost all the interstellar gas has been 
exhausted in the burst, is treated in the truncated spiral models with a burst
(TSB). 

In order to investigate if strong absorption Balmer lines necessarily imply a
burst, models of ``truncated spirals'' without burst have been also considered.
This case would correspond for example to the stripping of the interstellar gas
due to the intergalactic medium. 

Several types of spirals have been considered; the adopted SFR and metallicity,
as functions of the age have been derived from a standard chemical 
evolutionary model that includes an inflow and assumes the SFR to be
proportional to the gas fraction. 

This model provides the SFR and the metallicity as a function of time for
galaxies in the type range Sa-Sd. The model parameters for each Hubble type are
determined requiring the model SED to reproduce the observed colours and  gas
fraction  of local galaxies. A further model with constant SFR (flat model) has
been considered.

\subsection{Spirals with bursts}

Bursts have been added to models of Sa and Sc galaxies and to the flat model.
The explored range of the burst parameters is the same as in the case of
ellipticals. 

\begin{figure}
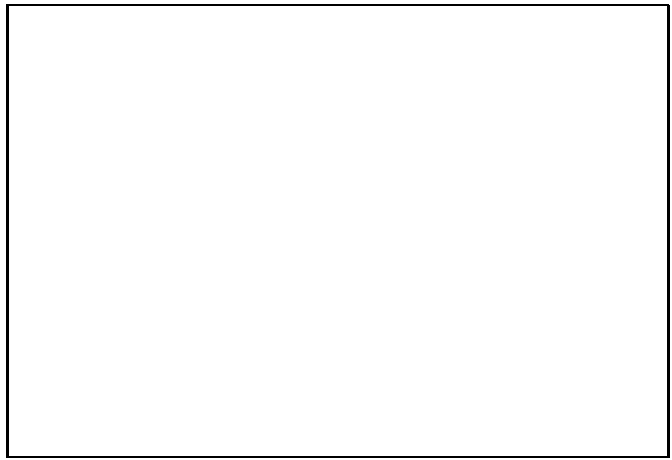

\picplace{6cm}
\caption[]{Time evolution ($\tau$ in Myr) of EW([OII]) 
(in \AA) after the
end of the burst in a Sa spiral for $\Delta g = 0.096$ (empty circles and
dotted line) and $\Delta g = 0.64$ (filled circles and full line). In these
models the SFR is not interrupted at the end of the burst}
\end{figure}

After the end of the burst such models still present a star formation process
and therefore show emission lines, although weaker. The equivalent width of the
 [OII]$\lambda$3727 line is plotted against $\tau$ in Fig. 6 for an Sa with a
burst in two cases: a burst with $\Delta g=0.096$ and  one with $\Delta
g=0.64$. The equivalent width quickly decreases at the end of the burst due to
the reduced SFR, reaching values lower than those characteristic of unperturbed
objects of the same morphological type ( $\approx \, 10 \,$ \AA).  This ensues
from the increase of the continuum around 3727 \AA $\,$ due to the presence of
a large number of young stars. EW([OII]) reaches a minimum for $\tau =30-100$
and afterwards, with a timescale depending on the intensity of the burst, it
tends to the unperturbed value. The maximum and the minimum values depend on
the intensity, the former increasing  and the latter decreasing with
$\psi_{1}/\psi_{0}$. 

The $\rm H \delta$ line, after the end of the burst, is in absorption, even if
the star formation is still active and moreover its intensity is particularly
high in the case of strong bursts. The presence of this feature together with 
[OII]$\lambda$3727  in emission can be interpreted in two ways: we are faced
with a galaxy having undergone a burst already stopped and with a residual
minor SFR or with a post-starburst elliptical in which the burst ended less
than 10 Myr ago. 

According to these models, spirals with bursts have blue  $\rm (B_J-R_F)$
colours ( $< 2$ mag) during  the whole evolution, with the exception of the
very early type spirals with weak bursts or stopped long ago. High values of
$\rm H \delta$ can be obtained without truncation, this however is necessary in
order to have spectra devoided of emission lines or red colours. 

\subsection{Truncated spirals with bursts}

Different results are obtained if, after the end of the burst, the SFR
completely ceases. Smaller percentages of mass involved in the burst are
required to reach intense absorption lines compared to the previous case. The
minimum $\Delta g$ required to have $\rm EW(H\delta)>3$ \AA $\,$ is however not
substantially different. For the rest the results are similar to the case of
bursts in ellipticals. The maximum equivalent width of the $\rm H\delta$ line,
typically reached at about 100 Myr from the burst end, is independent of the
underlying morphological type and depends only on $\Delta g$ (Fig. 4, filled
circles). 

\begin{figure}
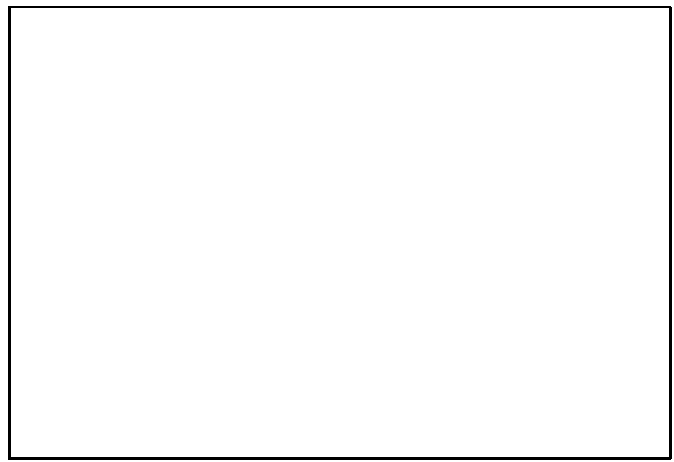

\picplace{6cm}
\caption[]{As Fig. 3 for two Sc galaxies 
truncated after the burst:
$\Delta g = 0.26$ (filled circles) and $\Delta g = 0.026$ (empty circles).
The empty and filled squares mark the positions of the unperturbed
elliptical and Sc models respectively}
\end{figure}

The time evolution in the  $\rm H\delta$-$\rm (B_{J}-R_{F})$ diagram shown in
Fig. 7, is similar to the corresponding diagram for ellipticals with bursts of
Fig. 3. As the evolutionary time scale is the same both in the elliptical and
the spiral burst models, the colour of the model is determined only by $\tau$.
The only difference is that, in the spiral case, the model does not relax to
the initial values but tends to the colours and equivalent width of the
unperturbed elliptical. 

\subsection{Truncated spirals}

Consider the case of a spiral in which the star formation has been completely
stopped at some instant. About 10 Myr later the spectrum does not show emission
lines at all. The colours evolve towards those of an elliptical with a
timescale depending on the morphological type and on the considered colour.
Fig. 8a shows the evolution of the $\rm (B_{J}-R_{F})$ colour for different
morphological types ($\tau=0$ marks now the end of the star formation process).
A different behaviour is presented by the (1550-V) (Fig. 8b) which becomes
redder than that of the elliptical. 

\begin{figure}
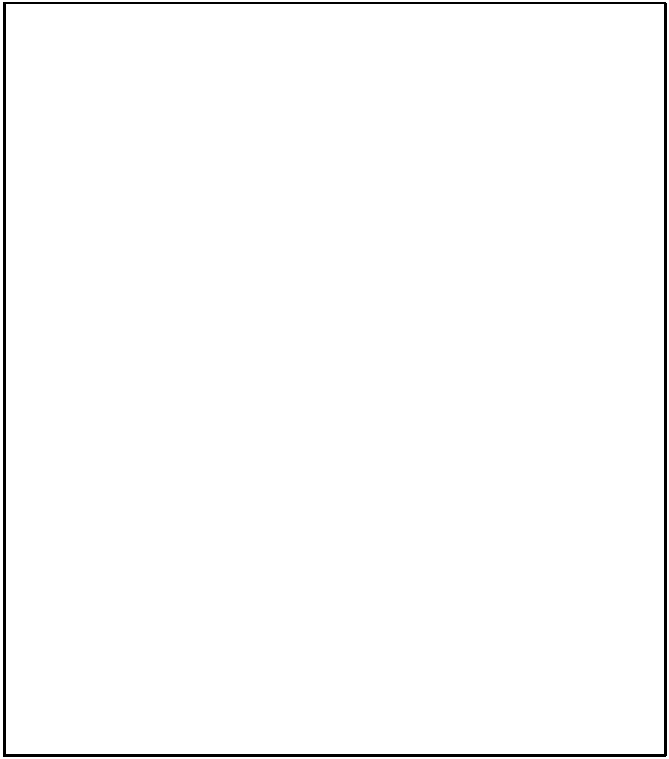

\picplace{10cm}
\caption[]{Truncated spirals without burst: time 
evolution ($\tau$ in
Myr) of the colour \bf a) \rm $\rm (B_J-R_F)$ and \bf b) \rm 
$\rm (1550-V)$  after the
end of the SF  for the Sa (open circles), Sc (filled circles) and constant SFR-
model (open squares)}
\end{figure}

A truncated spiral does not reach $\rm EW(H\delta) > 4.1 \,$ \AA, the maximum
value increasing with the type (2.3 for Sa, 3.6 for Sc and 4.1 for the flat
model). The truncation of the star formation alone therefore is unable to
interpret the high equivalent widths observed in distant clusters.

\section{A synthetic view} 

In this section we summarize the main general results derived from  our 
theoretical analysis.

\begin{figure}
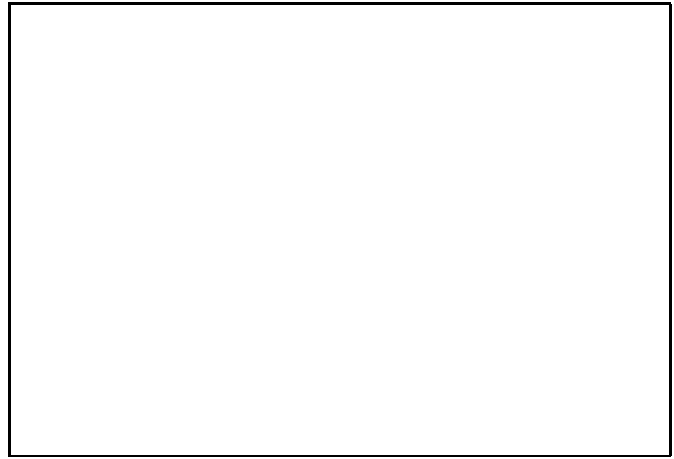

\picplace{6cm}
\caption[]{Post-starburst model spectra: elliptical with a burst
(solid line, $\Delta t = 500$ Myr, $\Delta g = 0.15$, $\tau = 500$ Myr);
truncated Sa with burst (short dashed line, 
$\Delta t =1$ Gyr, $\Delta g = 0.64$, $\tau = 700$ Myr); truncated Sc with
burst (long dashed line, $\Delta t =1$ Gyr, $\Delta g =0.26$, $\tau = 500$
Myr). The three models have the same $\rm EW(H\delta$) and the same $\rm 
(B_J-R_F)$ colour; fluxes are in arbitrary units}
\end{figure}

1) It has been seen that, concerning the diagrams of Fig. 3 and Fig. 7, the
same results are found independently of the original morphological type in
which the burst occurs. The question is therefore how to distinguish the
original galactic type. In Fig. 9 three  post-burst spectra are shown
corresponding to different types: E, Sa and Sc. All of them present the same
values of $\rm H \delta$ ( $\approx 4.5 \,$ \AA) and $\rm (B_J-R_F)$ ($\approx
2.1$). Only for $\lambda < 2000 \,$ \AA, do the spirals and the elliptical
significantly differ and consequently their (1550 -V) colours would allow their
distinction. Spirals are redder since they lack a population of metal-rich,
old, hot stars, present in ellipticals. This discrimination can be made only
when the bursts are old enough that galaxies are red, otherwise the UV spectral
region is dominated by the stars born during the burst. In fact the same models
of Fig. 9, viewed only 30 Myr after the burst end, show the same spectral
distribution also in the UV region. 

2) A current burst or one ended a few Myr ago is undoubtly witnessed by the
presence in the spectrum of an  $\rm H \delta$ line in emission or
emission-filled and a [OII]$\lambda$3727 emission line with extremely large
equivalent width (EW $\ge $ 25 \AA). This is however a sufficient but non
necessary condition for the presence of a burst.

3) EW($\rm H\delta$) reaches its maximum about 100 Myr after the end of the
burst and remains almost constant during the whole $\rm H \delta $-strong phase
(Fig. 3 and Fig. 7). Its value depends only on $\Delta g$ (see Fig. 4). As the
period in which EW($\rm H\delta$) is approximately at its maximum lasts long
enough, the probability of catching an object in this phase is high and this
allows the possibility of estimating $\Delta g$ provided that the following
conditions are met: the absence of the  [OII]$\lambda$3727  emission line, a
not too red colour ($\rm (B_{J}-R_{F}) < 2.1$) and of course an intense $\rm H
\delta$ absorption line. As the maximum of EW($\rm H\delta$) becomes saturated
at $\Delta g \simeq 0.20 $, for very strong bursts only a lower limit to
$\Delta g$ can be placed. Due to the steep rise in Fig. 4 the uncertainty in
the determination of $\Delta g$ for relatively weak bursts is large; the
accuracy becomes better for stronger bursts. 
 
\begin{figure}
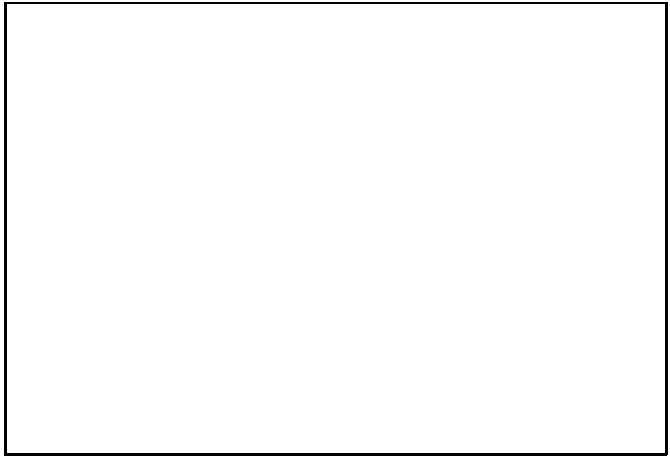

\picplace{6cm}
\caption[]{$\rm EW(H\delta)$-$\rm (B_J-R_F)$ diagram
divided into zones. Dashed lines delimit the region where the observed 
$\rm EW(H\delta)$ values of spirals from the Kennicutt's sample are placed}
\end{figure}

4) It is useful to analyze in detail the  $\rm H\delta$-$\rm (B_J-R_F)$ diagram
and to subdivide it into several areas, as shown in Fig. 10, according to the
partition into blue and red  and into  $\rm H\delta$-strong and $\rm
H\delta$-weak objects. The horizontal line at EW($\rm H\delta$)=1 \AA $\,$
together with the other full horizontal line defines the position of the
spirals (zone C). Zone F gives the position of red objects with very strong
$\rm H\delta$ lines. As shown in Fig. 3, an elliptical with an intense burst
will leave zone A, cross zones B when the burst is still active, C (rapidly in
a few million years), D, E or F (phase E+A) and will return again in A. 
In the case of a truncated spiral with burst, the starting point is located 
in region C, while the following evolution shows the same features as the 
ellipticals.

\begin{figure}
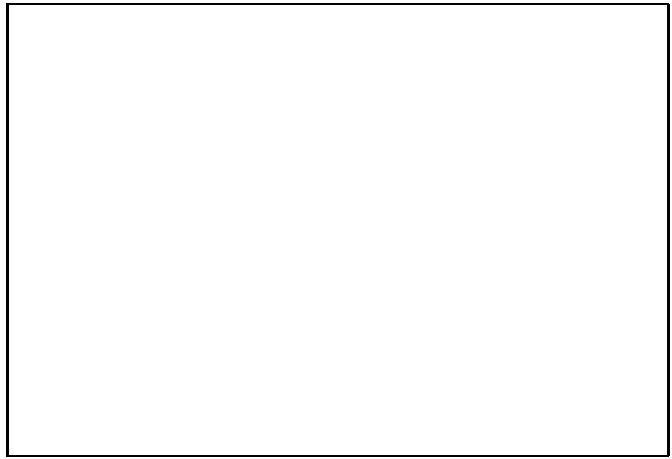

\picplace{6cm}
\caption[]{Location of the regions of Fig. 11 
in the parameter space
($\tau$ in Myr)}
\end{figure}

5) The position in the diagram of Fig. 10 allows some delimitation of the
parameters of the burst $(\tau$ and $\Delta g$), as it is clarified in Fig.
11, in which the location of the zones specified in the previous diagram is
presented. Physically zone B and zone D are not directly connected, because in
most cases a galaxy evolves through B, C and D areas. However the time spent in
area C is so short that this phase practically  cannot be distinguished in the
diagram. Moreover, due to the limited time resolution of each evolutionary
sequence, the boundary between zone B and zone D is uncertain, being located in
the interval 4 - 10 Myr. From the same figure we can conclude that, with the
exception of the weakest bursts, the evolutionary path crosses the same
regions; only the time spent in each area increases with $\Delta g$. 

\BTA

\caption[]{$\rm (B_J-R_F)$ colour ranges corresponding to the various $\tau$
during the $\rm H\delta$ strong phase}

\begin{flushleft}

\TAB{rrr}          
\hline
\noalign{\smallskip} 
$\tau$ & $3. < \rm H\delta < 4.5 (\AA)$  & $\rm H\delta>4.5  (\AA)$ \\
\noalign{\smallskip}
\hline
\noalign{\smallskip}
10  & 1.14-1.78 & 1.32-1.38 \\
30  & 1.41-1.98 & 1.54-1.61 \\
100 & 1.86-2.23 & 1.77-1.84 \\
500 & 2.12-2.21 & 2.03-2.16 \\
1000 &2.25-2.33 & ---    \\
2000 &   ---     & ---    \\
\noalign{\smallskip}
\hline
\ETAB

\end{flushleft}

\ETA

6) Provided that the $\rm H\delta$ line is strong, the time elapsed since the
end of the burst can be estimated from the $\rm (B_J-R_F)$  colour index. This
conclusion holds for any kind of underlying galaxy. Table 2 presents such
correlation. Any conclusion is of course influenced by the uncertainty
affecting the measure of the equivalent width; in the case of strong bursts the 
determination of $\tau$ is more precise. From all the models it turns out that 
1.5 Gyr after the end of the burst $\rm EW(H\delta)$ is always less than 3 \AA
$\,$ while 700 Myr after  $\rm EW(H\delta)$ is less than 4.5 \AA. This means
that, according to the models, no $\rm H \delta$-strong object can be redder
than $\rm(B_J-R_F)=2.35$ mag. In the following we shall see that this fact
creates some problems in the interpretation of some galaxies in clusters at
$\rm z=0.31$. 

\BTA

\caption[]{Minima percentages of galactic mass involved in the burst
required to detect such a  burst at some point of the evolution}

\begin{flushleft}

\TAB{ll}         

$\rm H\delta$ & 2.5 \\   
R-I &     0.5 \\
V-R  &   0.1 \\
$\rm B_J-R_F$ &  0.05 \\
$\rm D_{4000}$ &  0.01 \\
U-685 &  0.01 \\
$\rm H\alpha$ & 0.01 \\
B-V   &  0.01 \\
U-B   &  0.001 \\
OII   &  0.001 \\
1550-V & 0.0001 \\

\ETAB

\end{flushleft}

\ETA

7) Consider an elliptical galaxy to which a burst is added. We determined 
the minimum $\Delta g$ required to distinguish such a model from the
corresponding unperturbed one, when a given observable is considered and
the observational error is taken into account. Table 3 gives such values. 
In particular $\Delta g=0.025$ identifies the threshold between $\rm H
\delta$-strong and $\rm H \delta$-weak objects. $\Delta g=0.0005$ separates
blue and red objects in the $\rm (B_J-R_F)$ colour. It must be remarked that
the evolutionary phases and the time length during which the  various
quantities show a detectable difference with those of the unperturbed
elliptical, depend on the considered observable. Therefore if a given object
has a strong $\rm H \delta$ line, also the other quantities will show an
anomalous behaviour at some moment (not however necessarily the same). 
In particular this implies that an object in region E of Fig. 10 must have 
crossed in the past regions B and D. The same conclusion holds if the 
underlying galaxy is a spiral.

\section{Analysis of three clusters at $\rm z=0.31$}

In the following the previous theoretical analysis will be employed to
interpret the spectroscopic and photometric data of three clusters at $\rm
z=0.31$ (AC103, AC114 and AC118). CS87 have obtained spectra with a resolution
of 4 \AA $\,$ in the spectral region 4750-6750 \AA $\,$ (corresponding to
3625-5150 \AA $\,$ in the rest frame). Moreover they give the colour $\rm
(B_J-R_F)$, the apparent magnitude $\rm R_F$  and the equivalent widths in the
rest frame of the [OII]$\lambda$3727 and $\rm H\delta$ lines. 
CS87 confirmed spectroscopically the Butcher-Oemler effect detected
photometrically by Couch and Newell (1984). They also found that the red
peak of the colour distribution is about 0.2 mag bluer than the expected
location of the K-corrected unevolved ellipticals, a fact already noted in the
majority of the clusters sampled by Couch and Newell. This shift is larger than
the evolutionary correction expected from our models, unless the time scale
of star formation in ellipticals is considerably longer than 1.5 Gyr. 

\begin{figure}
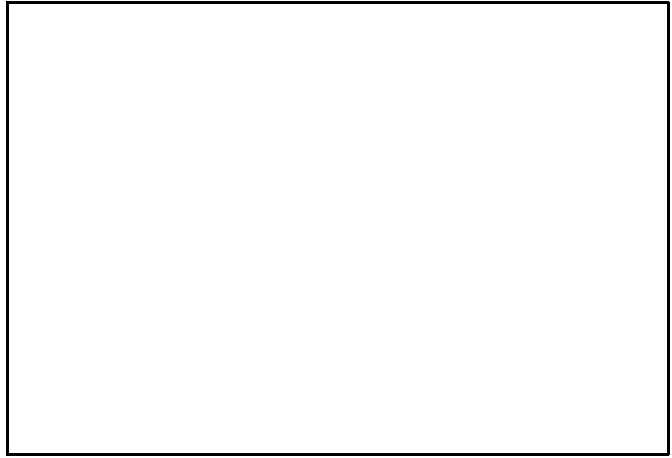

\picplace{6cm}
\caption[]{Observed quantities for the galaxies in the 
three CS87 clusters
(triangles). The equivalent width given is the rest frame value; the colour
is reddening corrected. Spiral models (Sa-Sd) are superimposed as filled 
circles}
\end{figure}

The equivalent width of  [OII]$\lambda$3727 of all the galaxies with $ \rm
EW([OII])>5 $ \AA $\,$ is plotted against  $\rm (B_J-R_F)$, corrected for
Galactic reddening (Fig. 12);  spiral models are also shown. The bluest objects
with the  strongest emission lines cannot be explained by models of normal
galaxies. 

Red galaxies have spectra very similar to those of local ellipticals. A
fraction of them, however, exhibits an intense absorption $\rm H\delta$ line
($\rm EW(H\delta)=3-5$ \AA), not found in the spectrum of normal ellipticals.
CS87 introduced three classes among blue galaxies: 1) with emission lines and
emission-filled Balmer lines; 2) with emission lines and moderate/strong 
absorption Balmer lines; 3) without emission lines and with strong absorption
Balmer lines. All types are present in each cluster, their frequency increases
with the class number (from  1 to 3). Most of the blue galaxies do not show
emission lines. Only one AGN has been found in AC118; the frequency of such
objects in these clusters agrees with that found in local clusters. 

CS87 interpreted red $\rm H\delta$-strong objects in terms of starbursts
occurred in ellipticals, while spiral progenitors were considered for types 1)
and 3). From the analysis of a set of absorption lines (H10(3795 \AA), H9(3840
\AA), H8(3890 \AA), CaIIK(3933 \AA), CN(4182 \AA) and the G band(4301 \AA)),
Jablonka \& Alloin (1995) concluded that all the red galaxies in the three
clusters are elliptical-like objects with traces of star formation 1-5 Gyr ago
and that 1/3 of the blue galaxies experienced a recent burst, while the
remaining blue objects could be interpreted as early-type spirals. 

CESS94 have also obtained further spectroscopy of AC114 and high spatial
resolution images with HST. The comparison of the four objects belonging to the
two samples demonstrates the uncertainty in the determination of the equivalent
widths. The new spectroscopic data do not introduce any further class, they
only increase the number of objects. The HST data of CESS94 provide information
about the morphological type and the presence of possible interactions or
mergers. 

The main HST results for AC114 can be summarized as follows:

1) the blue galaxies are mostly disk objects;

2) there is a high percentage of interactions/mergers among blue objects;

3) $\rm H \delta$-strong red galaxies are apparently isolated ellipticals.

\begin{figure}
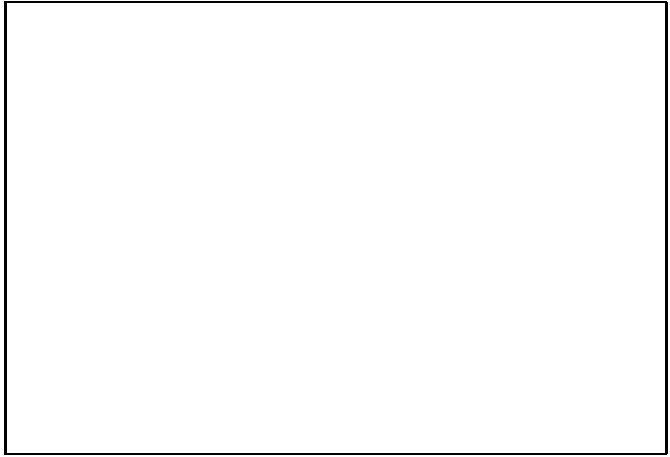

\picplace{6cm}
\caption[]{Observed data 
for the three clusters placed in the diagram
of Fig. 11. Circles are objects without emission lines (empty are blue, 
filled are red); crosses represent objects with a detectable [OII]$\lambda$3727
line. From right to left, filled squares show the position in the
diagram of the E, Sa, Sb, Sc and Sd models. The dotted lines show the 
range of $\rm EW (H\delta)$ measured from the spectra of the spiral
galaxies in Kennicutt's sample}
\end{figure}

Figure 13 presents the $\rm H \delta$ equivalent width and the $\rm
(B_{J}-R_{F})$ colour of the objects of the three clusters as well as the
subdivision into zones, previously described. Since the errors on the
equivalent widths are typically of 1 \AA $\,$ and those on the colour are of
0.1 mag, the position of the objects close to the boundaries need to be
considered with some caution. The E, Sa, Sb, Sc and Sd  models are also
plotted. While the colour range of our models agrees with the models used by
CS87 and the observations of Dressler and Gunn (1983), our model EW($\rm H
\delta$) values differ from those of CS87, the latter reaching higher values up
to 5 \AA $\,$ and increases steadily with the morphological type. 
A comparison of these models (Paper II) with the observations of local spirals
(Kennicutt 1992) confirms our computed values of EW($\rm H \delta$) which, for
the latest spirals, is the consequence of the emission-filling. When the
emission of the ionized gas is neglected our models agree with the results of
CS87. Both Kennicutt's observations and our models show that spiral galaxies
present the [OII]$\lambda$3727 emission line; therefore all the blue objects
without emission lines do not have the spectroscopic characteristics of normal
spirals. In particular for our spiral models with the adopted age is  $\rm
EW([OII])>10$ \AA. The presence of emission lines indicates the existence of
current or recent ($\tau < 10 \, \rm Myr$) star formation. Their absence
implies that in the last 10 Myr the star formation has involved less than 0.001
\% of the galactic mass. 

We analyze now the whole observational data set by the light of the different 
kinds of models, starting with ellipticals. 

{\it Ellipticals with bursts}

Figure 14a shows the same type of diagram of the previous figure in which,
instead of the observed data, the models of ellipticals with bursts are
located. In this figure models with emission $\rm H \delta$ line are not
included. These correspond to active bursts begun not long ago: the intensity
of the $\rm H \delta$ line in emission is large while the underlying continuum
is not enhanced yet.  Only one object has such a characteristic ( \#103 di
AC114, EW($\rm H\delta$)=-0.8). The limited number of such objects is amenable
to the quickness of the evolution or/and to the fact that burst time lengths
are typically longer. Moreover while the star formation is active the $\rm H
\delta$ remains weak. 

\begin{figure}
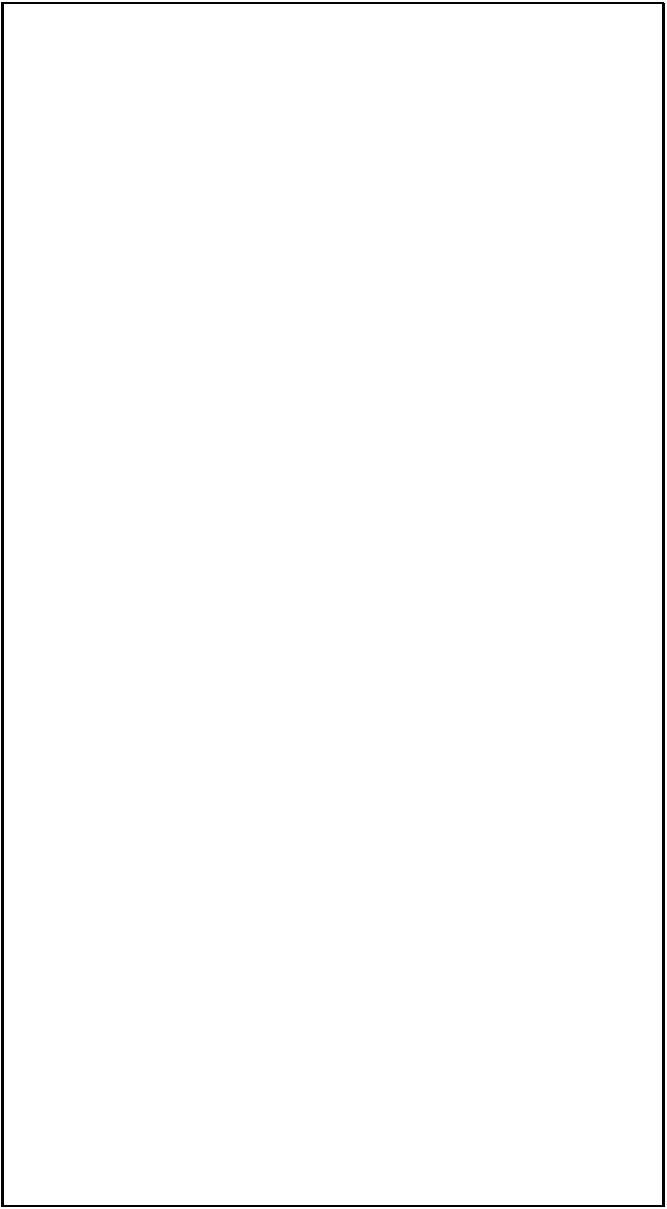

\picplace{16cm}
\caption[]{Results burst models 
are shown as circles (without emission
lines) and crosses (with $\rm EW([OII])>2$ \AA) for the whole burst parameter
space investigated. \bf a) \rm ellipticals with bursts; \bf b) \rm spirals with
bursts (without truncation); \bf c) \rm truncated spirals with bursts; \bf d)
\rm truncated spirals without bursts}
\end{figure}




Consider in detail the behaviour in the different areas.

-  {\bf zone A. } In this region the following models  are placed: normal 
ellipticals, ellipticals with a weak burst, ellipticals with an intense old 
burst ($\tau \geq 1-2 \, \rm Gyr$). 
The elliptical model has been computed with an average solar metallicity.
Reducing the metal content moves the model horizontally towards bluer colours.
Therefore in CS87's sample most objects in zone A could be interpreted as
ellipticals of various metallicity. 

Ellipticals with a weak burst and ellipticals with an intense old burst  can be
distinguished from the normal ellipticals on the basis of the (1550 - V)
colour, being respectively bluer and redder. Models with emission lines
represent ellipticals with a weak burst and $0 <\tau < 10$ Myr. This can be
understood remembering that a small amount of newly formed stars is able to
give rise to emission lines without modifying the colours (see Table 3). Only
one such a case is observed:  \#12 in AC103. 

- {\bf zone B. } In this zone are located galaxies with a current burst
consuming more than 0.05 \% of the total mass. Models present the  
[OII]$\lambda$3727 
line and the emission-filled $\rm H\delta$ line: this last feature requires
that the length of the burst is at least 100 Myr since otherwise the net  $\rm
H\delta$ flux would be in emission. Objects lacking emission lines do not admit
any interpretation on the basis of the models: they can be neither starbursts
nor post-starbursts. In the former case they should have emission lines, in the
latter they would have stronger absorption $\rm H \delta$. 

- {\bf zone C. } Normal spirals and ellipticals with bursts just ended $(\tau
\approx 10 \, \rm Myr$) lay in this zone: in the latter the  [OII]$\lambda$3727
 line is still
present while the emission filling in the $\rm H\delta$ line is decreasing. Only
one object falls in this region; this however is not ad odds with the number of
objects present in zone B, as the evolutionary time in zone C is very short.

- {\bf zone D. } Models appear with or without emission lines; both type of
objects are present among the observed galaxies. In order to reach this area a
 burst must have occurred with the following characteristics:  $\Delta g >
0.025$ and $\tau= 10-500$ Myr; for models with emission lines the time
elapsed since the burst end should not exceed a limit approximately equal to
30 Myr. It has been shown in Table 2 that in this phase the  $\rm (B_J-R_F)$
colour is correlated with $\tau$. 
Objects with $\rm EW(H \delta) \ge$ 7.5 \AA $\,$ are not expected due to the
saturation of $\rm H\delta$ with $\Delta g$ (see Fig. 4) and in fact only one 
galaxy in the CS87 sample exceeds this limit, within the observational error.

- {\bf zone E. } In this region the ``E+A'' galaxies are placed. These are 
post-starbursts with $\tau$ = 0.5 - 1.5 Gyr. In this area no object can be
found coming directly from zone A: all the $\rm H \delta$-strong galaxies must
have passed through a blue phase (see Table 3): more precisely it must
evolve through the path  B, C and D. 1-2 Gyr after the burst end, objects of
zone E evolve into region A, as it results from Fig. 3. 
Models exclude the presence of emission lines in E and in fact no such object 
is observed.

- {\bf zone F}. The same conclusions for zone E are valid. One aspect must be
stressed: the reddest ($\rm (B_J-R_F)>2.3$) objects with the strongest $\rm
H \delta$
line cannot be interpreted with the present models. Increasing $\Delta g$
affects the line in the right way but the colour becomes too blue. At least
four objects in the considered clusters have such characteristics.

To summarize, the whole observed scenario could be interpreted in terms of
starbursts in ellipticals, with the exception of a very limited number of 
objects, for which a further analysis is desirable.
 
We extend now our analysis by considering spiral models.

\it
Spirals with bursts.
\rm
Considering now models of spirals to which a burst is added, in Fig. 14b 
the $\rm EW(H\delta)$-$\rm(B_{J}-R_{F})$ diagram is presented with the usual
subdivision into zones. These models assume that the star formation is not 
interrupted after the end of the burst. They are characterized by 
emission lines and mostly by blue colours. This type of models is not able
to reproduce most of the observed galaxies of Fig. 14a both without emission
and E+A, however they could explain $\rm H\delta$-strong emission line objects.

\it
Truncated spirals with bursts.
\rm
This type of models  (Fig. 14c) explains the observed diagram as well as 
ellipticals with bursts: time scales and the covered zones are the same. It has
already been shown that, in case of a strong burst, the subsequent evolution is
driven by its characteristics and not by the original galaxy type. 

\it
Truncated spirals.
\rm
The maximum $\rm H \delta$ reached by this type of models is 4 \AA $\,$ in the
case of the latest types; moreover colours are never bluer than 1.7 mag (Fig.
14d). Therefore in order to explain the whole observed diagram, in particular
zones B and D,  a burst is necessary, being the sole truncation of the star
formation insufficient. Only a small fraction of the observations concerning
zone D could be interpreted by these models. Truncated spirals  cannot
interpret also objects in zone B without emission lines because they would
evolve above the spiral sequence.





\begin{figure}
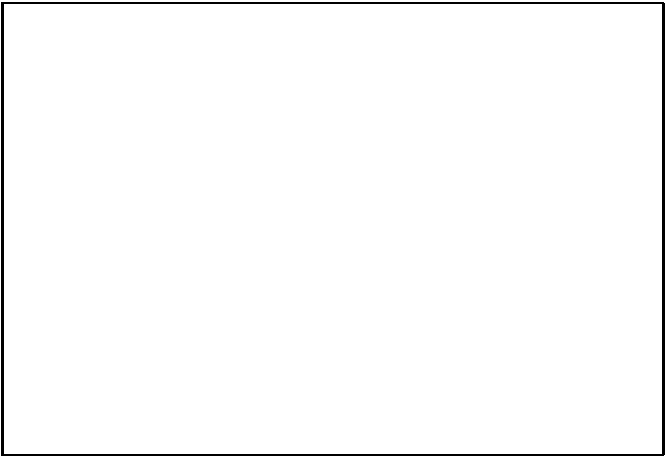

\picplace{6cm}
\caption[]{Distribution in $\Delta g$ for all 
the $\rm H\delta$-strong
galaxies with $\rm (B_J-R_F)< 2.3$ mag. Dashed lines show the same distribution
for blue objects only. The $\rm EW(H\delta)$ values correspondent to the
$\Delta g$ intervals are also shown}
\end{figure}

By putting together the results derived from all the kinds of models, the
histograms of $\Delta g$ and $\tau$ have been obtained for all the clusters
globally, as shown in Fig. 15 and Fig. 16.  In the first figure all $\rm H
\delta$-strong objects are included, whose colour is bluer than 2.3 mag. 
The values of $\Delta g$ are found from the theoretical relation between
EW($\rm H \delta)$ and  $\Delta g$ of Fig. 4. Let us consider as example a
galaxy with $\rm EW(H\delta)=4$ and $\rm (B_J-R_F)=2.1$. From Fig. 4 we can
conclude that $\Delta g = 0.05-0.10$ (therefore the fraction of galactic mass
involved is between 5 and 10 \%) and from Table 2, second column, a time $\tau$
between 100 and 500 Myr is derived. 

\begin{figure}
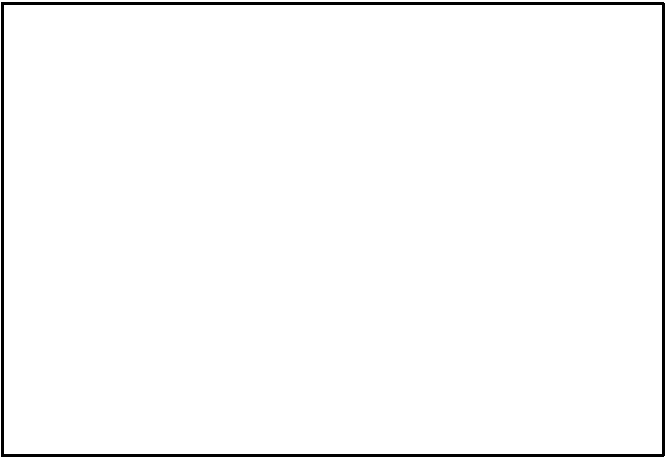

\picplace{6cm}
\caption[]{Distribution of $\tau$ for the galaxies in 
the three clusters:
shaded areas correspond to $\rm EW(H\delta) > 4.5$ \AA, empty areas to
$3 < \rm EW(H\delta) < 4.5$ \AA}
\end{figure}

The distribution of $\Delta g$ presents two peaks. When however the colour
distributions of each $\Delta g$ interval are analysed, it is found that most
of the cases corresponding to the small-$\Delta g$ peak are red objects and
therefore their $\rm H \delta$ is already decreasing. Neglecting the red
objects, the distribution has only the high-$\Delta g$ peak. We conclude that
the bursts identified in the CS87 sample involve in most cases a significant
fraction of the galactic mass, of the order of 30 \%. We stress that an
uncertainty of about 1 \AA $\,$ in  $\rm EW(H\delta)$ should always be
considered, accounting for the fact that the values obtained depend on the
method used to measure the equivalent width: comparing the values of $\rm
EW(H\delta)$ obtained from us and from Couch (private communication) of the
same spectra, we find differences inside the adopted uncertainty of 1 \AA. 

The histogram of $\tau$ (Fig. 16) is derived from the relation between this
timescale and the colour, shown in Table 2 in the case of strong $\rm H \delta$
line. Filled areas refer to objects with $\rm EW(H\delta) > 4.5$ \AA, empty
areas to  $\rm 3 < EW(H\delta) < 4.5 \,$ \AA. From a pure theoretical point of
view one would expect to observe in each interval a number of galaxies
proportional to the time interval. This is not the case since an excess of
cases with short $\tau$ is found; a possible interpretation could be the bias
towards current starbursts of the spectroscopic sample, due to the increased
luminosity during the burst. Galaxies whose estimated $\tau$ is greater than 1
Gyr are rare and this agrees with the relevant decreasing intensity of the $\rm
H\delta$ line found in the models.

\subsection{Analysis of the cluster Cl1358+6245 at $\rm z=0.33$}

The existence of the Butcher-Oemler effect has been questioned since clusters
of galaxies are in most cases selected in the optical and this could provoke a
contamination with background and foreground objects, even if the redshift is
determined spectroscopically (Koo 1988). 
A selection based on the luminosity in the X band should avoid this problem;
one of the clusters chosen with this criterium is Cl1358+6245 at $\rm z=0.33$,
studied by Fabricant et al. (1991). From the spectra they have derived $\rm
D_{4000}$, $\rm EW([OII])$ and \=H (average value of the equivalent widths of
$\rm H\beta$, $\rm H\gamma$ and $\rm H\delta$). Our elliptical model fits well
the observations of red passive galaxies (not showing any sign of recent star
formation). Models computed for post-starburst galaxies indicate that \=H, to a
first approximation, is equal to $\rm EW(H\delta)$, within the errors. The
analysis performed for the three AC clusters in terms of  $\rm EW(H\delta)$,
$\rm EW([OII])$ and $\rm (B_J-R_F)$ has been  repeated for this cluster by
using $\rm D_{4000}$ instead of the colour, since the equivalence between these
two quantities has been previously demonstrated. 

From the comparison observations-models for this cluster, we conclude that the
same characteristics of the anomalous objects found in the AC clusters are
recovered also in  Cl1358+6245 even if the selection criteria are completely
different. 

\section{Summary}

In this paper a spectrophotometric model has been presented which reproduces
the emission of both the stellar and the gaseous component of a galaxy. This
model is particularly helpful in understanding phenomena in which a burst of
star formation is involved. This is the case of most of the galaxies of the
three clusters observed  by CS87, in which the presence of a burst is the 
only way of interpreting the observations.

It has been shown that both the elliptical and the spiral can be the original
morphological type, contrary to the findings of CS87. When the spiral type is
adopted as the underlying galaxy, in addition to the burst also the truncation
is required to give an overall picture of the observations. This however does
not exclude that, for a minority of objects, also other mechanisms (truncated
spirals, spirals with bursts) can be effective. 

The knowledge of the three quantities $\rm EW([OII])$, $\rm EW(H\delta)$ and
the colour $\rm (B_J-R_F)$) allows one to determine, in several cases, 
the burst characteristics. The largest number of constraints is
found for the $\rm H \delta$-strong objects, in which the time elapsed since
the burst end ($\tau$) and the galactic mass fraction of gas exhausted in it
($\Delta g$) can be estimated. 
The histograms of $\Delta g$ and $\tau$ have been obtained; they show that the
bursts involve in most cases a significant fraction of the galactic mass
(typically 30 \% or more). 

However the mentioned observed quantities do not uniquely define the original
morphological type. Our previous analysis has shown that such a distinction can
be obtained by using  an UV colour such as (1550 -V), which up to now has not
been observed in these clusters. Even the direct morphology obtained with HST
in principle is not able to identify the original type due to the short
dynamical time connected to possible changes in morphology.

\begin{acknowledgements}
We would like to thank A.~Arag\'on-Salamanca and H.R.~Butcher for useful 
discussions and comments.
We are also grateful to R.~Kurucz for sending his
new stellar atmosphere models, to M.~Auddino who supplied his spiral models
and to M.~Radovich for running the programme Cloudy for us. We acknowledge the
availability of the Kennicutt's galaxy atlas and the Jacoby et al's stellar
library from the NDSS-DCA Astronomical Data Center.
\end{acknowledgements}

\end{document}